\documentclass{icrc29}
\usepackage[english]{babel}
\usepackage[latin1]{inputenc}
\usepackage{times}
\usepackage[T1]{fontenc}
\usepackage{hyperref}
\usepackage{epsfig}
\usepackage{graphicx}
\usepackage{amssymb}

\hypersetup{
pdftitle={AUGER internal note},
pdfauthor={Antoine Letessier-Selvon, CBPF - In2p3/CNRS},
pdfborder={0 0 0},
pdfview={Fit}
}
		  
\begin{document}
\title[Excess maps, source flux upper limits] %
{A procedure to produce excess, probability and significance maps and to compute point-sources 
flux upper limits}

\author[Billoir, Letessier-Selvon]%
{P. Billoir (Billoir@in2p3.fr) \& A. Letessier-Selvon (Antoine.Letessier-Selvon@in2p3.fr)}

%


\presenter{}

\maketitle

\begin{abstract}
A short note to propose a procedure to construct excess maps, probability maps 
and to calculate point source flux upper limits.
\end{abstract} 

\section[Introduction]{Excess maps}
In the following we do not discuss how the coverage map is constructed, be it with shuffling, 
with the semi-analytical method or whatever. 
we simply assume that we have, in the case of a perfectly isotropic CR sky, a  prediction for the average number of
cosmic ray we should observe from a given direction on the sky.
Let $M_{bg}(\Omega)$ be this coverage function and $M_{bg}(k)$
its value integrated over the pixel k (centered on direction $\Omega_k$) of a pixelized representation of the sky.    

Given a set of events and their individual pointing accuracy we can construct
a CR density map $M(k)$ just by counting the number of events that fall in pixel $k$. We can also construct 
a smooth CR density map distributing those events over the map according to their pointing accuracy. 
Let $M_s$ be such a smooth function and $M_s(k)$ its value integrated over pixel k.

We can construct excess map or relative excess map as :
\begin{enumerate}
\item $ M_e(k) = M_s(k) - M_{bg}(k);$ or
\item $ M_{re}(k) = M_s(k)/M_{bg}(k).$ 
\end{enumerate}
Since we cannot predict from the exposure (we do not know the flux) the total number of events expected in 
case of an isotropic sky we normalize $M_{bg}$ to the total number of events observed ($N = \int M_s(\Omega) d\Omega)$.
It can be useful to filter (convolute) the above maps with a global Gaussian point spread function of the type 
$P(\Omega,\theta_0^2)d\Omega = e^{-\theta^2/(2\theta_0^2)}d\Omega$ where $\theta$ is the angular distance to the center
of the pixel we are filtering, to emphasize (or to enhance)  structures that may be present at a given scale $\theta_0$. 

\begin{figure}[t]
\begin{minipage}[t]{0.5\textwidth}
\begin{center}
\epsfig{file=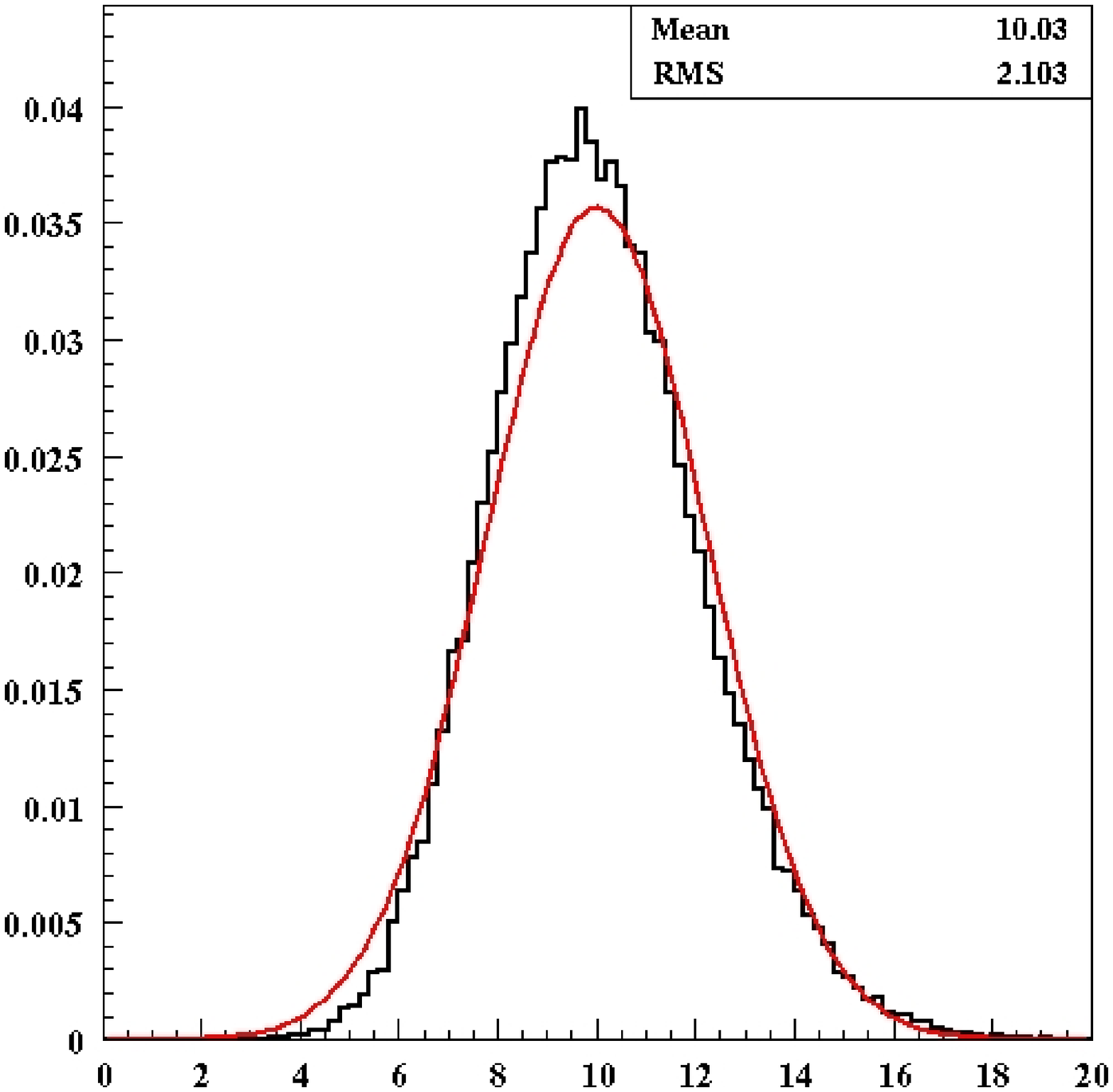,width=0.97\textwidth}  
\end{center}
\end{minipage}
\begin{minipage}[t]{0.5\textwidth}
\begin{center}
\epsfig{file=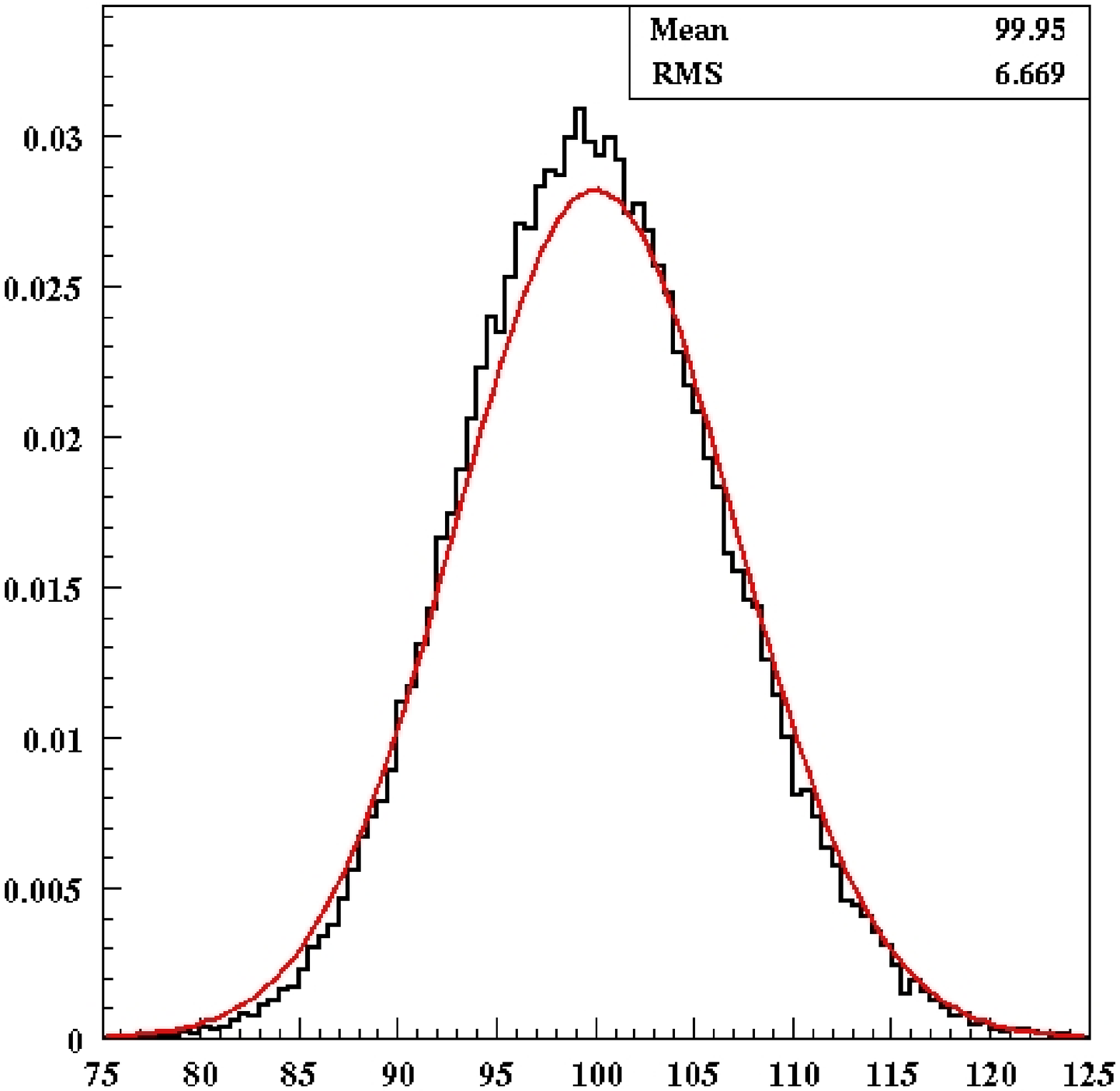,width=1.0\textwidth}  
\end{center}
\end{minipage}
\caption{True background distribution after Gaussian filtering on a 1.5$^\circ$ scale (size of filtering does not
matter up to at least 15$^\circ$). 
The average number of events (N$\varepsilon_0$) in a 1.5$^\circ$ disk is 5 in the left plot and 50 in the right one. 
The overlayed function is the approximated normal distribution (${\cal N}(2N\varepsilon_0, N\varepsilon_0)$) with parameters as predicted by Eq.~\ref{eq:meanX} and Eq.~\ref{eq:varX}.}
\label{fig:1}
\end{figure}

\section[Probability maps]{Probability maps}
In a probability map, each pixel carries the probability that 
a uniform sky leads to a larger number of event observed in that pixel than what we have measured. Without
any filtering and if we do not distribute events on the sky according to their pointing accuracy, constructing
such a map is straightforward as the above probability is given by Poisson statistics.
\begin{equation}
M_{prob}(k) = P(n>M(k); M_{bg}(k))
\label{eq:probMap}
\end{equation}

If we have done filtering  we can either use a Monte Carlo
technique to estimate the above probability or approximate the filtered random variable distribution with an appropriate 
normal distribution whose parameters are derived below in the case of a Gaussian filtering. 
\subsection{MC calculation}
For the MC technique we can construct $N$ smoothed density map $M^{mc}_s$ in exactly the same
way as we have constructed our signal density map $M_s$. The corresponding probability map will then be given by
\begin{equation}
M_{prob}(k) = \frac{\mbox{number of maps satisfying: } \{M^{mc}_s(k) > M_s(k)\}}{N}
\label{eq:MCProbMap}
\end{equation}
\subsection{Gauss approximation}
For Gaussian filtering with parameter $\theta_0<<1$, such that we can work on the plane in a disk of radius $\Theta$  ignoring
the contribution of the points outside of this disk and the variation of the coverage on that disk, we can model the 
distribution of the expected background. Let $\mu = N/(\pi\Theta^2)$ be the uniform CR background density then 
the average weight $<w>$ of a background event is :
\begin{equation}
<w> = \frac{1}{\pi\Theta^2}{2\pi}\int_0^\Theta \theta e^{-\theta^2/(2\theta_0^2)}d\theta = 2\left (\frac{\theta_0}{\Theta}\right )^2 = 2\varepsilon_0
\label{eq:meanX}
\end{equation}
The above integral is still accurate to a few percent for $\theta_0$ as large as $ 15^\circ$ and $\Theta=60^\circ$.

While the variance of the weights is given by :
\begin{equation}
V[w] = <w^2> - <w>^2 = \varepsilon_0(1 - 4\varepsilon_0) \simeq \varepsilon_0
\label{eq:varX}
\end{equation}
since we assume that $\varepsilon_0 << 1$.

For N events the random variable {\bf X}=$\sum_i^N w_i$ has an average <{\bf X}> =$ 2N\varepsilon_0$ and a variance V[{\bf X}] = $N\varepsilon_0$.
The probability density function of {\bf X} can be adequately modeled (especially on the excess side) by a normal law 
${\cal N}(2N\varepsilon_0, N\varepsilon_0)$ even for an average background count in a circle of radius 
$\varepsilon_0$ ($N\varepsilon_0$) as low as 5. See Figure 1. 

The probability map is then obtained from the filtered signal and coverage map as :
\begin{equation}
M_{prob}(k) = \int_{M_s(k)}^{+\infty} {\mathcal N}(x; M^{mc}_s(k), M^{mc}_s(k)/2) dx
\label{eq:GaussProbMap}
\end{equation}
Or equivalently a significance map can be obtained as :
\begin{equation}
M_{sig}(k) = \frac{M_s(k) - M^{mc}_s(k)}{\sqrt{M^{mc}_s(k)/2}}
\label{eq:GaussSigMap}
\end{equation}

\underline{\bf Remark :} If the number of pixels in the maps are very large so that the probability to have an event falling
in a given pixel is much smaller than 1 then, without filtering,  the probability map gets very difficult to interpret
(at least visually) as it will be only composed of pixel with values 0 or 1.

\section[Point source fluxes]{Optimizing point source detection}\label{optim}
Here we want to address the question of how to optimize the search region to detect an eventual
point source located at $\Omega_0$. For simplicity we rotate our coordinate system so that
$\Omega_0 = 0$.  In the following we assume that the aperture of the array does not vary significantly
on the scale of the detector angular resolution ($\sigma$), therefore the background can be considered uniform 
in the region we are looking at. 

The average number of events from a uniform CR distribution expected from that direction is given by :
\begin{equation}
\mu_{bg} = \int_{\Delta E} \mathcal{A}(E) \Phi_{CR}(E)\int_{\Delta \Omega} W(\Omega)d\Omega dE
\label{eq:nbg}
\end{equation}

where $\Phi_{CR}(E)$ is the uniform CR flux per unit solid angle , time, surface and energy.
$\mathcal{A}(E)$ is the aperture (per unit time, surface, and energy) which we assumed independent 
of $\Omega$, and $W(\Omega)$ is the weight function we want to optimize to detect point-like sources.

From an hypothetical point source $s$ in that same direction the average number of events is given by :
\begin{equation}
\mu_s = \int_{\Delta E} \mathcal{A}(E) \Phi_s(E)\int_{\Delta \Omega} W(\Omega)\eta(\Omega)d\Omega dE
\label{eq:ns}
\end{equation}

where $\Phi_s(E)$ is now the source flux per unit time surface and energy and $\eta(\Omega)$ is the detector
point spread function (PSF), assumed independent of energy for simplicity.

The significance $S$ of the signal is given by the ratio of the signal from the source $n_s$ to the background 
fluctuation ($\sqrt{V[n_{bg}]}$). To maximize $S$ we need to maximize this ratio at each energy that is to maximize :
\begin{equation}
R(W) = \frac {\int_{\Delta \Omega} W(\Omega)\eta(\Omega)d\Omega}{\sqrt{V\left [ \int_{\Delta \Omega} W(\Omega)d\Omega\right ] }} 
\label{eq:Smax}
\end{equation}
with respect to $W(\Omega)$.

A general result of signal processing theory tells us that the optimal choice 
for $W$ is the detector response function, i.e. taking  $W$ proportional to  $\eta(\Omega)$ in our case. 
With such a choice and using  $W(\Omega) = e^{-\theta^2/(2\sigma^2)}$ the $R$-ratio becomes :

\begin{equation}
R_{Gauss} = \frac {\pi\sigma^2\Phi_s(E)}{\sqrt{\pi\sigma^2\Phi_{CR}(E)}} = \sigma\Phi_s(E)\sqrt{\frac{\pi}{\Phi_{CR}(E)}} 
\label{eq:Rgauss}
\end{equation}
where we have used Eq.~\ref{eq:varX} for the variance of the background.

It has been argued that a top-hat weight function $W$ can also maximize this ratio. In case of a top-hat weighting function
$W_\alpha(\Omega) = H(\theta-\beta\sigma)$ where $H$ is the Heavyside function and $\beta$ the distance in units of 
$\sigma$ up to which we want to extend the source integration, we have :
\begin{equation}
R_{\beta} = \frac {2\pi\sigma^2\Phi_s(E)(1-e^{-\beta^2/2})}{\sqrt{\pi\beta^2\sigma^2\Phi_{CR}(E)}} = 2\left (\frac{1-e^{-\beta^2/2}}{\beta}\right ) R_{Gauss} 
\label{eq:RTH}
\end{equation}
$R_\beta$ reaches a maximum value of $0.9\times R_{Gauss}$ for $\beta^2 = e^{\beta^2/2} - 1$ (i.e. $\beta$ = 1.585). Which shows
that a top-hat weighting function is always less efficient than weighting with the detector PSF. 

We studied the evolution of the R-ratio in the case where the PSF used for filtering has a different parameter than
the true PSF of the detector. Calling $\alpha$ the ratio between the filter parameter and the true PSF parameter 
Eq.~\ref{eq:Rgauss} and Eq,~\ref{eq:RTH} become~:

\begin{equation}
R_{Gauss}^\alpha = \frac {2\alpha}{1+\alpha^2}R_{Gauss}^1\mbox{\rm~~~~and~~~~}
R_\beta^\alpha = 2\left (\frac{1-e^{-(\alpha\beta)^2/2}}{\alpha\beta}\right )R_{Gauss}^1
\end{equation}

\begin{figure}[t]
\label{fig:alpha}
\begin{center}
\epsfig{file=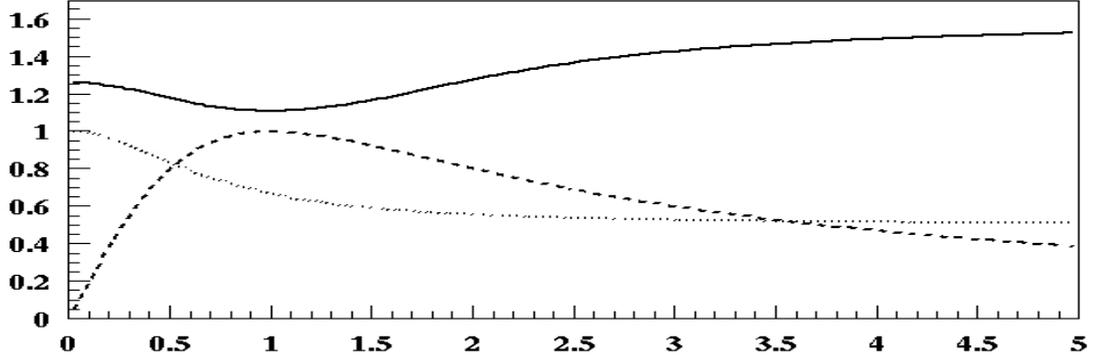,width=0.97\textwidth, height=5cm}  
\end{center}
\caption{Ratio of $R_{Gauss}^\alpha$ to $R_\beta^\alpha$ (solid-line), $R_{Gauss}^\alpha$ 
(dotted-line) and V[$n_s$]/<$n_s$> as a function of $\alpha$.}
\end{figure}

We show on  Figure~\ref{fig:alpha} the evolution of $R_{Gauss}^\alpha$ as a function of alpha as well the ratio of 
$R_{Gauss}^\alpha$ to $R_\beta$ for  $\beta = 1.585\sigma$. We see that for all $\alpha$ $R_{Gauss}^\alpha$ is larger than 
$R_\beta^\alpha$ and, as expected, that $R_{Gauss}^\alpha$  is maximum for $\alpha=1.0$.

\section{Flux upper limits}
To obtain flux upper limits at a certain confidence level (CL) one needs first to derive from the number of events observed
and the expected background contribution an upper limit, at the same CL, on the source contribution to the observation.

\subsection{Top-Hat weighting functions}
If we use a top-hat weighting function to count our events around a certain direction then the source contribution upper 
limit at CL $\alpha$, in that window can be directly derived using Poisson statistics. 
We call $\mu^\alpha_s$ this upper limit and we have\footnote{Formally the variable $\mu^\alpha_s$, $n_{bg}$ and $n_{bg}$ should all carry the letters $TH$ to mark the fact they are all computed using a Top-Hat window filter. However, 
For simplicity we dropped this symbol.}:

$$
\mu^\alpha_s = \mu_\alpha - n_{bg}
$$
where $n_{bg}$ is the expected number of count in our top-hat window and $\mu_\alpha = \mu_s^\alpha+n_{bg}$ is such that the Poisson cumulative 
probability of observing more  than $n_{obs}$ events in the window is larger than $\alpha$ :
\begin{equation}
P(n>n_{obs}; \mu_\alpha) = \sum_{i=n_{obs}+1}^{+\infty}P(i;\mu_\alpha)  > \alpha. 
\label{eq:uplimPoiss}
\end{equation}
or, equivalently such that the probability to observe at most $n_{obs}$ events is less than $1-\alpha$~:
\begin{equation}
P(n<=n_{obs}; \mu_\alpha) = \sum_{i=0}^{n_{obs}}P(i;\mu_\alpha)  < 1-\alpha. 
\end{equation}
This formula can be solved numerically or can be approximated for $\mu_{bg}$ sufficiently large 
(above about 10 for CL in the range [0.8-0.95], larger CL would require larger average background) 
using a Gaussian distribution ($\mathcal{N}(n_{bg},n_{bg})$ for the background contribution. 
This leads to :
\begin{equation}
\mu^\alpha_s = (2n_{obs}+C_\alpha^2 + \sqrt{4n_{obs}C_\alpha^2+C_\alpha^4})/2 - n_{bg}
\label{eq:uplimTH}
\end{equation}
where $C_\alpha$ is such that $\int_{-\infty}^{C_\alpha}\mathcal{N}(x;0,1)dx = \alpha$ ($C_\alpha=1.64$ for $\alpha$=95\%).

\subsection{Gauss weighting functions}
In this case the event count around a given direction does not follow a Poisson distribution (it is no longer integer) 
so we can either rely on a Monte Carlo evaluation of $\mu^\alpha_s$, throwing background events uniformly around the 
source direction with an average background count equal to our background expectation in the filter window 
and throwing source events with a distribution proportional to  $e^{-\theta^2/(2\sigma^2)}$ finding the source normalization 
for which in CL\% of the case the MC count is larger than the observed count\footnote{Here also the variables should 
carry the letter $G$ to remind they are now computed with a Gaussian filter but again we have dropped this superscipt for clarity.}.

Alternatively we can model the distribution of the sum of the background and signal count and derive the upper limit 
$\mu_s^\alpha$ from this distribution. For the background we have already shown that the count distribution can be 
approximated by a normal law $\mathcal{N}(n_{bg}, n_{bg}/2)$ a similar calculation gives for the signal weights :
\begin{equation}
<w> = \frac{1}{2\pi\sigma^2}{2\pi}\int e^{-\theta^2/(\sigma^2)}\theta d\theta = \frac{1}{2}
\end{equation}
and
\begin{equation}
<w^2> = \frac{1}{2\pi\sigma^2}{2\pi}\int e^{-3\theta^2/(2\sigma^2)}\theta d\theta = \frac{1}{3}
\end{equation} 
For a source of average intensity $\nu_s$ our event count is $n_s = \sum_{i=1}^n w_i$ where $n$ is distributed  according to 
a Poisson law of average $\nu_s$ so our average count becomes :
\begin{equation}
\mu_s = <n_s> = \sum_n \mathcal{P}(n;\nu_s) \sum_{i=1}^n <w_i> = \frac{\nu_s}{2}
\end{equation}
with variance :  
\begin{equation}
V[n_s] = \sum_n \mathcal{P}(n;\nu_s) <\sum_{i=1}^n w_i \sum_{j=1}^n w_j> - (\nu_s<w>)^2 = 
\frac{\nu_s}{3} = \frac{2\mu_s}{3}
\label{eq:vns}
\end{equation}

Using for the distribution of the total count a Gaussian approximation $\mathcal{N}(n_{bg}+\mu_s, n_{bg}/2+2\mu_s/3)$ 
the upper limit of CL $\alpha$ on the source count is given by the solution of
\begin{equation}
n_{obs} - (n_{bg}+\mu_s) = C_\alpha\sqrt{\frac{n_{bg}}{2}+\frac{2\mu_s}{3}}
\label{eq:uplimG}
\end{equation}
which can easily be solved analytically.

On Figure~\ref{fig:alpha} we show the evolution of the ratio $V[n_s]/\mu_s$ as a function of the $\alpha$ parameter introduced in 
section~\ref{optim} $V[n_s]/\mu_s(\alpha) = (1+\alpha^2)/(1+2\alpha^2)$. We see that for $\alpha > 1$ Eq.~\ref{eq:vns} is an upper bound on $V[n_s]$ and will therefore 
lead to conservative upper  limit if we mistakenly use  a filter with a larger $\sigma$ parameter than the detector PSF.

In Table~1 we compare the 95\% CL upper limit obtained from Eq.~\ref{eq:uplimPoiss}, ~\ref{eq:uplimTH}
and~\ref{eq:uplimG} for various conditions of 
background and for an observed count equal to the background prediction. 
For the approximation of Eq.~\ref{eq:uplimG} we indicate the true CL of 
this upper limit as extracted from a Monte Carlo simulation.

\subsection{Fluxes}
If the source  spectral shape is the same as the overall CR spectral shape in the domain $\Delta E$
where we want to place a flux upper limit, that is if $\Phi_s(E) \propto \Phi_{CR}(E).$ then the aperture 
part of the integral contribute in the same way for the background and for the signal and we can relate the source 
flux upper limit to the ratio of $\mu_s^\alpha$ to our expected background in the following way :
\begin{equation}
\Phi_s^\alpha = \frac{\beta^2}{1-e^{-\beta^2/2}}\frac{\mu_s^\alpha(TH)\Phi_{CR}\pi\sigma^2}{n_{bg}(TH)} = \frac{3.5\mu_s^\alpha(TH)\Phi_{CR}\pi\sigma^2}{n_{bg}(TH)}
\end{equation}
for an optimal top-hat window of width $\beta =1.585$ in units of $\sigma$ the detector Gaussian PSF parameter, 
and, for a Gauss filter of parameter $\sigma$. 
\begin{equation}
\Phi_s^\alpha = \frac{4\mu_s^\alpha(G)\Phi_{CR}\pi\sigma^2}{n_{bg}(G)}
\end{equation}
where we have explicitely indicated that the values of the source count upper-limit and of the expected background 
count depend on the filter type used.  

Given the values presented in Table~1, the upper limits from the two methods only vary by a factor 
\begin{equation}
f = (4\beta^2\mu_s^\alpha(G)/(3.5\mu_s^\alpha(TH)
\end{equation}
which is approximately 15\% giving a small advantage to the Gauss filter.
We should however keep in mind that this is the best case comparison for the top-hat window (1.585$\sigma$) 
as any other value (possible if one does not know exactly the detector resolution) would increase the difference.

\begin{table}[!t]
\label{tab:uplim}
\begin{center}
\begin{tabular}{c|c|c|c|c|c|}
$n_{bg}$ (in a 1-$\sigma$ disk) & 2.5 & 5.0 & 10.0 & 25.0 & 50.0  \\ \hline 
Top-Hat window (exact)&  5.6 & 8.1 & 9.8 & 14.9 & 20.5\\ \hline 
Top-Hat window (approx.)&  5.3 (0.94)  & 7.8 (0.94)  & 9.5 (0.95) & 14.6 (0.95) & 20.2 (0.95) \\ \hline 
Gauss filter (exact)& 3.4 & 4.4 & 5.7 & 8.7 & 12.2  \\ \hline
Gauss filter (approx.)& 3.6 (0.96) & 4.7 (0.96) & 6.2 (0.96) & 9.1 (0.96) & 12.5 (0.96) 
\end{tabular}
\caption{(95\% CL upper limit on the source count for an optimal top-hat window and a Gaussian filter with the detector PSF 
parameter. In all case the observed number of events was taken to be equal to the background expectation (or the nearest 
interger if needed). The exact line corresponds to the exact solution given by the numerical solution or an MC integration. 
The approx. line corresponds to the Gauss approximation with its corresponding CL (as computed from a MC) in parenthesis,}
\end{center}
\end{table}

\section{Conclusions}
We have shown that using a Gauss filter to produce sky maps or to compute flux upper limits is an optimal choice
to enhance the sky feature at a particular scale. However, the gain in sensitivity is rather small (15\%) compared 
to a top-hat filter, if the detector resolution is properly know. In other cases the improvement offered by Gauss filtering 
is always larger.  For situation where the angular resolution varies from event to event, we have shown that 
using the largest value give good results and safe upper limits. It is however better, and all the above considerations
are still valid,  to adapt the filter on an event by event basis. That is, for each event entering the count, to filter it 
with a Gauss filter matching  its resolution. 
\end{document}